\documentclass[12pt]{article}

\usepackage[fleqn]{amsmath}
\usepackage{amssymb} % e.g. \gtrsim
\usepackage{hhline}
\usepackage[scale={.75,.75}]{geometry}
\usepackage[T1]{fontenc}
\usepackage{bbm} % BlackBoard letters
\usepackage{fleqn} % equations are not centered
\usepackage{mathrsfs} % e.g. nice Lagrange-L with \mathscr{L}
\usepackage{longtable}
\usepackage{caption}
\usepackage{cite}
\usepackage{epsfig}

\newcommand{\U}[1]{\ensuremath{\text{U}\!\left(#1\right)}}
\numberwithin{equation}{section}
\numberwithin{table}{section}
\allowdisplaybreaks

\begin{document}
\date{\mbox{ }}

\title{ 
{\normalsize     
DESY 09-026\hfill\mbox{}\\
September 2009\hfill\mbox{}\\}
\vspace{1cm}
\bf Light Moduli in \\Almost No-Scale Models\\[8mm]}
%
%\vspace{2cm} 
\author{Wilfried~Buchm\"uller, Jan M\"oller and Jonas~Schmidt\\[2mm]
{\normalsize\it Deutsches Elektronen-Synchrotron DESY, Hamburg, Germany}
}
\maketitle

\thispagestyle{empty}

\begin{abstract}
\noindent
We discuss the stabilization of the compact dimension for a class of  
five-dimensional orbifold supergravity models. Supersymmetry is broken by the 
superpotential on a boundary.
Classically, the size $L$ of the fifth dimension
is undetermined, with or without supersymmetry breaking, and the effective
potential is of no-scale type. The size $L$ is fixed by quantum corrections to the K\"ahler potential,
the Casimir energy and Fayet-Iliopoulos (FI) terms localized at the boundaries. 
For an FI scale of order $M_\mathrm{GUT}$,
as in heterotic string compactifications with anomalous $\U1$ symmetries,
one obtains $L \sim 1/M_\mathrm{GUT}$. A small mass is predicted for the scalar
fluctuation associated with the fifth dimension, 
$m_{\rho} \lesssim m_{3/2}/(L M)$.
\end{abstract}

\newpage

\maketitle

\section{\label{sec:intro} Introduction}

Higher-dimensional supergravity and superstring theories provide a promising 
framework for the unification of matter, gauge interactions and gravity 
\cite{wit85}. These theories possess vacua with unbroken supersymmetry and 
flat higher-dimensional Minkowski space. It is a challenging task 
to find four-dimensional non-supersymmetric locally stable Minkowski or
de Sitter vacua, with compact extra dimensions smaller than the 
electroweak scale. 

Classically, size and shape of the compact dimensions are generically 
undetermined. Stabilization can occur as a result of quantum corrections.
In field theory, these include the Casimir energy \cite{ac83,pp01,ghx05,
gh05,bcs08} and localized Fayet-Iliopoulos (FI) terms \cite{lnz03}. 
Loop- and $\alpha'$-corrections play a crucial role for the stabilization 
of volume moduli 
in string theory \cite{bbx02,bbx05,bhk05,ccq08}, in addition to fluxes. 
Furthermore, in string theory and field theory nonperturbative 
corrections to the superpotential are often required to achieve a complete 
stabilization of the compact dimensions \cite{kkx03,ls00}.

Recently, it has been suggested that the interplay of Casimir energy
and localized Fayet-Iliopoulos terms can lead to the stabilization of 
the compact dimensions
\cite{bcs08}. For a FI mass scale ${\cal O}(M_\mathrm{GUT})$, as it occurs in 
some compactifications of the heterotic string \cite{bls07}, one then obtains
for the size of the compact dimensions $L \sim 1/M_\mathrm{GUT}$. The height
of the barrier which separates four-dimensional from ten-dimensional
Minkowski space is $\mathcal{O}(m_{3/2}^2M_{\mathrm{GUT}})$. It therefore
vanishes for unbroken supersymmetry.

In this paper we study the interplay of supersymmetry breaking and FI terms in supergravity.
 We shall consider the simplest case of 
five-dimensional orbifold models, which include the dynamics of 
the radion superfield (cf.~\cite{ls00,mp01,pst05}). Such models can be 
considered a toy version for anisotropic compactifications of
ten-dimensional string theories, which have one `large' compact dimension. 
Due to the no-scale structure of the K\"ahler potential, it is impossible 
to realize non-supersymmetric locally stable Minkowski
or de Sitter vacua at tree level \cite{bda04,gs06,cgx08}. 
The radion flat direction needs to be lifted by quantum 
corrections to the K\"ahler potential, which always include the Casimir energy.

As we shall see, perturbative corrections to the K\"ahler potential, together with a non-zero brane superpotential, imply `almost no-scale' models, similar to the one proposed by Luty and Okada \cite{lo03}. The chiral superfield, which generates the expectation value of the superpotential,
couples to bulk fields. This coupling leads to a contribution to the radion potential,
which is of the same order of magnitude as the Casimir energy.
The resulting radion potential allows for metastable Minkowski or de Sitter vacua,
without the need of an additional `uplifting' mechanism. 

The paper is organized as follows. Section~2 describes the no-scale model
of a radion field coupled to a brane localized chiral superfield. The
general structure of `almost no-scale' models is
analyzed in Section~3, where also a formula for the radion mass is derived.
The stabilization induced
by localized FI terms is worked out in Section~4, which is followed
by a brief summary in Section~5. 

\section{\label{sec:model} A class of no-scale models}

Consider the bosonic part of the action of five-dimensional (5D)
$\mathcal{N}=1$ supergravity compactified on $S_1/\mathbb{Z}_2$,
with  bulk and brane contributions
\begin{equation}
S_5 = S_{\textrm{bulk}} + \delta(y)S_{\textrm{vis}} 
+ \delta(y-L)S_{\textrm{hid}}\ ,\notag
\end{equation}
where
\begin{equation}
S_{\textrm{bulk}}= \frac{M_5^3}{2} \int d^4x\, \int_0^L dy \sqrt{-g_5} 
\left\{ R_5 - \frac{1}{2}H^{MN}H_{MN} + \mathcal{L}_{\textrm{bulk}}\right\}\ .
\end{equation}
Here $H_{MN}=\partial_M B_N - \partial_N B_M$ is the field strength of 
the graviphoton, the spin-1 component of the supergravity multiplet.
Dimensional reduction of this action on the background metric 
\begin{equation}
ds_5^2=g_{\mu\nu}(x) dx^{\mu}dx^{\nu} + r^2(x) dy^2\,
\end{equation}
leads to
\begin{equation}
S_4=\frac{M^2}{2} \int d^4x \sqrt{-g} r 
\left\{ R - \frac{1}{r^2}\partial_{\mu} B_5 \partial^{\mu} B_5 
+ \mathcal{L}^{(4)}_{\textrm{bulk}} \right\}
+ S_{\textrm{branes}}[g^{\mu\nu}]\ ,\label{JF}
\end{equation}
where we have only kept $g_{\mu\nu},g_{55}$ and $B_5,$ which have
even $\mathbb{Z}_2$ parity. The remaining fields $g_{\mu5}, B_{\mu}$ are $\mathbbm Z_2$ odd and thus do not have light modes.
$M = \sqrt{M_5^3 L}$ is the 4D Planck mass, for a stabilized radion with $r_0 =1$ in the vacuum. Note that the radion field, i.e., the scale factor of the fifth dimension,
is dimensionless and has no quadratic kinetic term. Due to the bulk-brane structure, $r$ couples 
non-universally to the matter sector, hence it is not a Brans-Dicke scalar.

After a conformal transformation of the metric, 
$g_{\mu\nu}\rightarrow r^{-1}g_{\mu\nu}$,
one finds for the action in the Einstein frame,
\begin{align}
S_4=\frac{M^2}{2} \int d^4x \sqrt{-g}
&\left\{ R - \frac{3}{2r^2} g^{\mu\nu}\partial_{\mu} r \partial_{\nu} r 
- \frac{1}{r^2}\partial_{\mu} B_5 \partial^{\mu} B_5 \right. \nonumber\\ 
& \left.\hspace*{1cm} + 
\frac{1}{r}\mathcal{L}^{(4)}_{\textrm{bulk}}[rg^{\mu\nu}] \right\}
+ S_{\textrm{branes}}[rg^{\mu\nu}]\ .\label{EF}
\end{align}
This action contains a quadratic kinetic term for the radion field. 
Note the presence of the unusual factor $\,3$, which
will reappear in the K\"ahler potential below. This factor indicates that the kinetic term is solely
generated by the conformal transformation.
 
A globally supersymmetric theory is characterized by a holomorphic superpotential
$W(z)$ and a real function $\Omega(z,\bar{z})$ which yields the
kinetic terms 
\begin{equation}\label{eq:Lkinglobal}
\mathcal{L}^\textrm{global}_{\textrm{kin}} = 
\Omega_{i\bar{j}}\partial_{\mu}z^i\partial^{\mu}z^{\bar{j}} \ , \quad
\Omega_{i\bar{j}} \equiv \partial_i\partial_{\bar{j}}\Omega\ . 
\end{equation}
In the corresponding supergravity theory kinetic terms and scalar potential are
determined by the K\"ahler potential
\begin{equation}
K=-3M^2 \ln\left(-\frac{\Omega}{3 M^2}\right) \ ,
\end{equation}
with
\begin{equation} \label{eq:Lkin}
\mathcal{L}_{\textrm{kin}}^{\rm local} = 
K_{i\bar{j}}\partial_{\mu}z^i\partial^{\mu}\bar{z}^{\bar{j}}
\end{equation}
and
\begin{equation}
V_F = e^{K/M^2}\left[(W_i + 
M^{-2}K_i W)K^{i\bar{j}}(\bar{W}_{\bar{j}}
+ M^{-2}K_{\bar{j}}\bar{W})-3M^{-2}|W|^2 \right]\ .
\end{equation}

Let us now consider a model with minimal field content and include
one brane chiral superfield $X$ with canonical kinetic term, such that
\begin{equation}
\Omega = -\frac{3M^2}{2}\left(T+\bar{T}\right) + X\bar{X}\  .
\end{equation}
The Einstein frame component action (\ref{EF}) is then obtained for the K\"ahler potential 
\cite{f05}
\begin{equation}
K=-3M^2\ln\left(\frac{T+\bar{T}}{2}-\frac{X\bar{X}}{3M^2}\right)\ .\label{kp}
\end{equation}
The scalar component of the radion superfield contains the brane field $X$,
\begin{equation}
T=r+\frac{X\bar{X}}{3M^2}+i\sqrt{\frac{2}{3}}B_5\ ,
\end{equation}
compensating for the non-diagonal entries in the K\"ahler metric.

\begin{figure*}[t]
\begin{center}
\includegraphics[height=6.2cm]{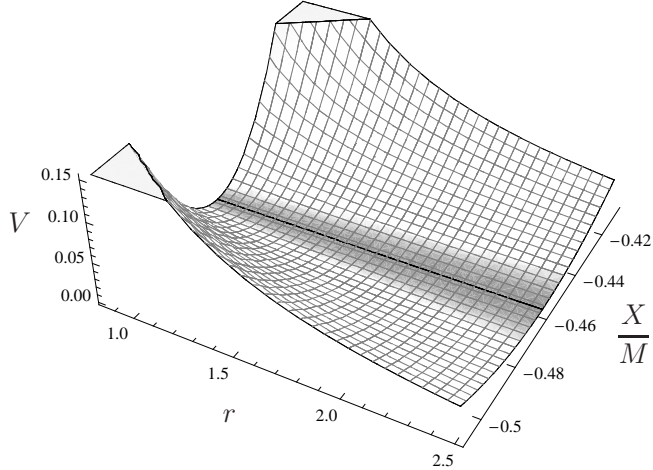}
\caption{\label{fig:toypot}Example of a no-scale potential, in units of $m^2_{3/2}M^2$. It follows from the quadratic superpotential (\ref{eq:W}) for the choice $\sqrt{\sigma_0}\simeq0.46$.}
\end{center}
\end{figure*}

The K\"ahler potential (\ref{kp}) has no-scale structure \cite{knx8384},
\begin{equation}
K^iK_i=3 M^2\ ,
\end{equation}
which is characteristic for a universal K\"ahler modulus.
Hence, the negative-definite contribution to the scalar potential vanishes, 
and one obtains
\begin{equation}\label{treepot}
V_F =  \frac{1}{r^2} W_X\bar{W}_{\bar{X}} \ . 
\end{equation}
The equations of motion
\begin{equation}
\partial_r V_F = 0\, , \quad \partial_X V_F = 0\ ,
\end{equation}
are simultaneously satisfied at stationary points of the superpotential,
\begin{equation}
\partial_X W|_{X_0} = 0\, .
\end{equation}
The potential then vanishes for all values of $r$, satisfying the Minkowski 
condition $V_F=0$, and the size of the 
compact dimension is undetermined (cf.~Fig.~\ref{fig:toypot}).

The K\"ahler potential does not depend on $B_5$, the imaginary part of the 
complex scalar $T$. At minima of the superpotential $W$ also the radion
is a flat direction. Hence, the corresponding two scalar masses vanish,
\begin{equation}
M^2_1 =0\ ,\quad M^2_2 = 0\ ,
\end{equation}
whereas the masses of real and imaginary part of $X$ are equal and positive,
\begin{equation}
M^2_3  = M^2_4 = \frac{1}{4} W_{XX}\bar{W}_{\bar{X}\bar{X}} \  .
\end{equation}
In the limit $|W_{XX}| \rightarrow \infty$, the degrees of freedom of 
$X$ decouple, and it becomes a spectator field, i.e., only its vacuum
expectation value (VEV) is relevant.

For non-vanishing superpotential, supersymmetry is spontaneously broken. 
The fermionic component of the radion superfield acts as the goldstino. 
The gravitino mass is given by
\begin{equation}
m_{3/2}^2 = e^{K/M^2} \frac{|W|^2}{M^4}
= r^{-3} \frac{|W|^2}{M^4}\  .
\end{equation}
As expected in no-scale models, the gravitino mass `slides' with the 
expectation value of the radion field.
 
The potential depicted in Fig.~\ref{fig:toypot} illustrates the continuous 
vacuum degeneracy which is generic for no-scale models. It is 
well known that K\"ahler potentials of the type \mbox{$K=-3 M^2 {\rm ln}\, r$} do 
not admit non-supersymmetric Minkowski vacua with a positive 
definite mass matrix \cite{bda04,gs06,cgx08}. 
A necessary condition for the latter can be formulated as \cite{gs06} 
\begin{align}
R_{i\bar j k \bar l}G^{\bar l}G^k G^{\bar j} G^i &<6 M^2\ ,
\end{align}
where $R_{i\bar j k \bar l}$ denotes the Riemann curvature of the K\"ahler manifold, and
\begin{align}
G&=K+M^2 {\rm ln} \frac{|W|^2}{M^6}\ .
\end{align}
The scalar potential is then given by
\begin{align}
V&=m_{3/2}^2 \left( G^iG_i-3 M^2 \right)\ .
\end{align}
For the two-field no-scale K\"ahler potential (\ref{kp}) vanishing of the 
vacuum energy implies
\begin{align}
R_{i\bar j k \bar l}G^{\bar l}G^k G^{\bar j} G^i &=6 M^2\ .
\end{align}
This result holds for any superpotential $W(X,T)$, even in the presence of nonperturbative corrections.
Therefore at least one flat direction is unavoidable.\footnote{Note that this 
argument also applies to the mechanism of \cite{ls00}, where the $F$-term uplift
induces a flat direction in the hidden sector.} We conclude that loop corrections are crucial for the stabilization of the radion in a Minkowski vacuum.

\section{Almost no-scale models} 

Quantum corrections change the real function $\Omega_0$ of no-scale models to
\begin{equation}
\Omega = \Omega_0 + \Delta \Omega\ ,
\end{equation}
where
\begin{equation}\label{kappa}
\Omega_0 = -3M^2 \left(\frac{T+\bar{T}}{2} - \frac{\sigma}{3}\right)\  ,\quad
\Delta\Omega \equiv 3M^2 r\kappa(r,\sigma)\  , \quad \sigma \equiv \frac{X\bar{X}}{M^2}\  .
\end{equation}
The corresponding K\"ahler potential is given by
\begin{align}
K&=-3M^2\ln\left[-\Omega_0\left(1+\frac{\Delta\Omega}{3M^2\Omega_0}\right)\right]\nonumber\\ 
&=-3M^2\left[\ln\left(\frac{T+\bar{T}}{2}-\frac{\sigma}{3}\right)
	+\ln(1-\kappa)\right]\  .
\end{align}
In the following we shall analyze the effect of the correction $\kappa$, which
turns the no-scale model of the previous section into an almost no-scale
model.

It is straightforward to calculate the ${\cal O}(\kappa)$ correction to
the no-scale potential,
\begin{align}\label{potential}
V_F = &\frac{1}{r^2}\,W_X\bar{W}_{\bar{X}}( 1 + 2\kappa 
+ \partial_r\left(r\kappa) 
- 3r \partial_\sigma \left(\sigma 
\partial_{\sigma} \kappa\right)\right) \nonumber\\
&+ \frac{3\left(X W_X \bar{W} + W \bar{X} \bar{W}_{\bar{X}}\right)}{M^2 r^2}
\partial_r\left(r\partial_{\sigma}\kappa\right)
- \frac{3 W\bar{W}}{M^2 r^2}\left(2\partial_r \kappa 
+ r \partial_r^2\kappa\right) \  .
\end{align}
The tree level minimum $X_0$ is now shifted to $X_0 + \Delta X$.
At linear order in $\Delta X$, the extremum condition 
\begin{align}
\left. \partial_X V_F \right|_{X_0+\Delta X}=0
\end{align}
 implies
\begin{equation}
\label{eq:deltaXres}
\Delta X = \frac{3W}{M^2W_{XX}}\left( 
-\bar{X}\partial_r\left(r\partial_{\sigma}\kappa\right)
+\frac{\bar{W}}{M^2W_{XX}}X\left(2\partial_r\partial_{\sigma}\kappa
+r\partial_r^2\partial_{\sigma}\kappa\right)\right)\Big|_{X_0,r_0}\  .
\end{equation}
Our systematic expansion in $\kappa$ is consistent as long as $|\Delta X|/|X_0|
\leq {\cal O}(\kappa)$. According to Eq.~(\ref{eq:deltaXres}) this holds if 
$|W_{XX}| \geq {\cal O}(|W|/M^2)$, i.e.,
\begin{equation}\label{xmass}
M_{3,4} \geq {\cal O}(m_{3/2})\  .
\end{equation}
Note that the corresponding fermion mass has to satisfy the same bound. 

The resulting leading order effective potential can then be read off from 
Eq.~(\ref{potential}),
\begin{equation}\label{radpot}
V^{(1)}(r,\sigma) =
- \frac{3|W|^2}{M^2} \left(\frac{2}{r^2}\partial_r\kappa(r,\sigma) 
+ \frac{1}{r}\partial^2_r\kappa(r,\sigma)\right)  \  .
\end{equation}

The stabilization of the radion at $r_0$ leads to a mass term for the 
corresponding scalar fluctuations,
\begin{equation}
r=r_0 + \delta r= 1+\sqrt{\frac{2}{3}}\rho\ , 
\end{equation}
where the definition of $\rho$ renders a canonical kinetic term.
The mass matrix of the complex scalars $T$ and $X$ has the eigenvalues
\begin{align}
M^2_1&=0\, ,\quad M^2_2 = \frac{|W|^2}{M^4}
\left(4\partial_r^3\kappa + \partial_r^4\kappa\right)\Big|_{X_0,r_0}
+\mathcal{O}(\kappa^2 m_{3/2}^2) \ , \label{mrad}\\
M^2_3 &=M^2_4 =\frac{1}{4}W_{XX}\bar{W}_{\bar{X}\bar{X}}\Big|_{X_0} 
+ \mathcal{O}(\kappa m_{3/2}^2)\  .
\end{align}
We conclude that the radion mass is $\mathcal{O}(\kappa)$ relative to the 
gravitino mass.\footnote{In the case of $\alpha'$-corrections, 
a similar relation for the radion mass has been obtained in \cite{cgx08}.} 
Note that this result does not depend on details of the 
stabilization mechanism. It is unavoidable whenever the vacuum is 
stabilized  by quantum corrections to the K\"ahler potential, which can
be treated perturbatively.

\section{\label{sec:ans}Perturbative stabilization of the radion}

In the previous section we discussed how quantum corrections deform the 
no-scale K\"ahler potential such that a stable, non-supersymmetric Minkowski 
vacuum can emerge. We shall now present a specific example where the quantum
corrections leading to Casimir energy and localized Fayet-Iliopoulos
terms are taken into account. In terms of these corrections the size $L$ of 
the extra dimension can be explicitly calculated.

In general, there is a contribution to $\kappa$ from the Casimir energy of 
the gravitational multiplet \cite{rss03} and other massless bulk fields, 
\begin{equation}
\Delta\Omega_{\textrm{C}}(r)= -\frac{1}{2 L^2}\left(A r^3 + 3 B r^2 
+\frac{C}{r^2}\right)
\equiv 3M^2r\kappa_{\textrm{C}}(r)\  ,
\end{equation}
which, according to (\ref{radpot}), corresponds to the potential
\begin{equation}\label{casimir}
V^{(1)}_{\rm C}(r) =
\frac{3|W|^2}{M^4 L^2 r^2}\left(Ar + B + \frac{C}{r^4}\right) \  .
\end{equation}
The Casimir energy (\ref{casimir}) vanishes for $W = 0$, i.e., for
unbroken supersymmetry. The constant $C$ is determined by the number of 
massless degrees of freedom in the bulk, the constants $A$ and $B$ are bulk 
and brane tensions, respectively. They are needed for the renormalization of 
the divergent Casimir energy and depend on the renormalization scale 
(cf.~\cite{ghx05,bcs08}).
These constants have been used to stabilize the radion at a minimum with
vanishing cosmological constant \cite{pp01}.\footnote{Note, however, that the
choice of the constants has to be consistent with supersymmetry \cite{bb03}.}
Our expansion around no-scale models is 
consistent as long as $A$ and $B$ are $\mathcal{O}(\kappa)$. For simplicity,
we choose $A=B=0$ in the following. As we shall see, radion stabilization in 
a Minkowski vacuum can still be achieved by fine tuning the remaining parameters of 
the scalar potential. 

In addition, massive bulk fields contribute to the Casimir energy. The resulting 
term in the effective radion potential is known to take the form \cite{pp01,lo03}
\begin{align}\label{eq:Vmass}
	V^{(1)}_{\rm C'}(r) 	=\left. \frac{3|W|^2}{M^4 L^2 r^2}
\frac{C'}{r^4}\right( &  \frac{M^2_\textrm{bulk} L^2 r^2}{3}\textrm{Li}_1\left(e^{-M_\textrm{bulk} L\, r}\right)
\nonumber \\ & \bigg.
	+M_\textrm{bulk} L\, r\,\textrm{Li}_2\left(e^{-M_\textrm{bulk} L\, r}\right)
	+\textrm{Li}_3\left(e^{-M_\textrm{bulk} L\, r}\right)\bigg)\  ,
\end{align}  
with the polylogarithmic functions 
\begin{equation}
\textrm{Li}_s\left(e^{-M_\textrm{bulk} L\,  r}\right)\equiv\sum_{k=1}^{\infty}\frac{e^{-k M_\textrm{bulk} L\,  r}}{k^s}\  .
\end{equation}
The constant $C'$ in (\ref{eq:Vmass}) is related to the number of degrees of freedom with mass $M_\textrm{bulk}$,
and will be specified below.
 Note that  
 $\kappa_{\rm C'}(r)$ can be obtained by integrating Eq.~(\ref{radpot}) for the potential (\ref{eq:Vmass})  (cf.~\cite{rss03}), 
which, however, is not required for our further calculations.

There are further corrections to the potential in the 
presence of brane-localized kinetic terms. Their contribution corresponds to 
a two-loop effect \cite{gh05, gh08} and is therefore subleading. 
Moreover, in string theory the K\"ahler potential is modified by supersymmetric 
loop corrections and $\alpha^{\prime}$-corrections,
which could be treated as additional contributions to the function $\kappa$.

In orbifold compactifications, generically
Fayet-Iliopoulos terms of anomalous U(1) symmetries arise at fixed points 
\cite{lnz03,bls07}. They induce a non-trivial vacuum configuration of the 
scalar sector: Bulk fields that are charged under the U(1) symmetry develop 
vacuum expectation values and become massive.
These VEVs ensure vanishing $F$- and $D$-terms in the bulk and at the
fixed points. In the simplest case of one hypermultiplet, containing the $\mathcal{N}=1$ chiral multiplets $H$ and $H^c$, one has
\begin{align}
\Delta\Omega_{\textrm{bulk}}&=H\bar{H}+H^c\bar{H^c}\  ,\\
\Delta\Omega_{\textrm{brane}}&=\frac{\lambda'}{M_5^3}
\left(H\bar{H}+H^c\bar{H^c}\right)X\bar{X}\  .
\end{align}
A detailed analysis \cite{ahk03} shows that if the sum of the FI terms is 
non-zero, one of the two chiral multiplets, say $H$, develops an $r$-dependent 
VEV, while $\langle H^c \rangle=0$. In the 4D theory one then obtains 
(cf.~(\ref{kappa}))
\begin{align}
\Delta\Omega_{\rm FI} &= \int^L_0 dy \left[r\langle H\bar{H}\rangle +
\delta(y-L)\,  \frac{\lambda'}{M^3_5}\langle H\bar{H}\rangle\, X\bar{X}\right]
\nonumber\\
&=\xi\left(1+\frac{\lambda X\bar{X}}{M^2 r}\right)
\equiv 3M^2 r \kappa_{\rm FI}(r,\sigma)\  .
\end{align}
Here $\xi$ is the sum of the two FI terms localized at the fixed points 
at $y=0$ and $y=L$, and $M_5^3 L = M^2$, provided $r_0 = 1$. The different
couplings $\lambda$ and $\lambda'$ reflect the discrepancy  between the 
condensate at $y=L$ and its average value. The function $\kappa_{\rm FI}$ 
corresponds to the effective radion potential (cf.~(\ref{radpot})) 
\begin{equation}\label{fi}
V^{(1)}_{\rm FI}(r,\sigma) =
-\frac{2  \lambda \sigma  }{r^3} \frac{\xi|W|^2}{M^4} \  .
\end{equation}
Note that the $r$-dependent background field value results in a deformation of the Kaluza-Klein spectrum. The special case $\xi=0$, accompanied by strong localization of the bulk fields, was 
discussed in \cite{gqr04}. Here we consider nearly constant VEVs. We then expect that the backreaction on the internal geometry remains negligible, such that the flat orbi\-fold is a valid approximation. However, small warping could be treated as an additional contribution to the $\kappa$ correction (cf.~\cite{gh08} and references therein).

\begin{figure}
\begin{center}
\includegraphics{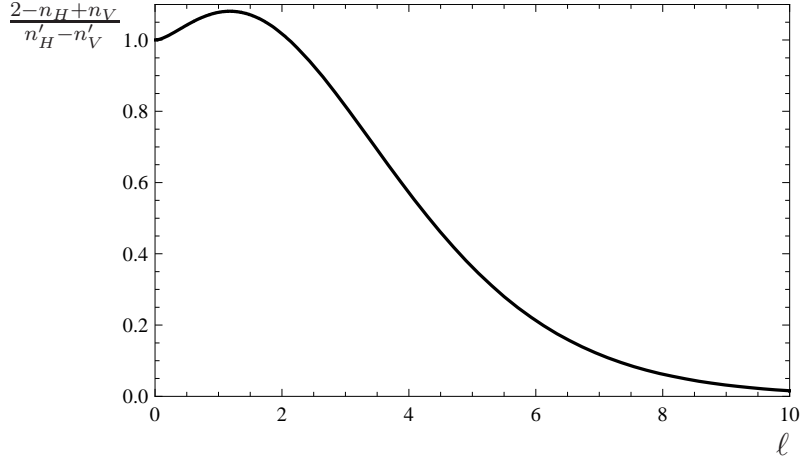} 
\caption{\label{fig:ratiom} The relation between the bulk field content and
the size of the compact dimension $L=\ell/M_\textrm{bulk}$. The plotted 
ratio of multiplicities has a maximum at $\hat{\ell}\simeq1.2$. 
}
\end{center}
\end{figure}

Furthermore, the VEV $\langle H\rangle$ breaks the anomalous U(1) and the
corresponding gauge boson acquires a mass $M_V = \mathcal{O} (\sqrt{\xi})$, 
like the hyperscalars. For simplicity, we assume a common mass parameter 
for the U(1) vector- and massive hypermultiplets.\footnote{Their contribution 
to the Casimir energy was neglected in \cite{bcs08}.} $\mathcal{O}(1)$
mass differences would not change our results qualitatively. With 
$\xi=\mathcal{O}(M_{\rm GUT}^2)$ (cf.\cite{bls07}), one has
\begin{equation}\label{susymass}
M_H = M_V = M_{\textrm{bulk}} = \mathcal{O} (M_{\rm GUT})\  .
\end{equation}
In terms of the dimensionless parameter $\ell$ defined by
\begin{equation}\label{size}
L = \frac{\ell}{M_{\textrm{bulk}}}\  ,
\end{equation}
the resulting radion effective potential reads up to terms ${\cal O} (\kappa)$, 
\begin{align}\label{pot}
V^{(1)}(r,\sigma) &=
V^{(1)}_{\rm FI}(r,\sigma)+{V^{(1)}_{\rm C}}(r)+V^{(1)}_{\rm C'}(r)
\notag\\
	&=\frac{3|W|^2}{M^2 r^2}\frac{M^2_{\rm bulk}}{M^2}
	\left[-\frac{2 \lambda \sigma}{3 r^3}\frac{\xi}{M^2_{\rm bulk}}
	+\frac{C}{\ell^2 r^4}\right.\notag\\&\left.\qquad\qquad\quad
	+\frac{C'}{\ell^2 r^4}\left(\frac{\ell^2r^2}{3}\textrm{Li}_1\left(e^{-\ell r}\right)
	+\ell r\textrm{Li}_2\left(e^{-\ell r}\right)+\textrm{Li}_3\left(e^{-\ell r}\right)\right)\right]\  .
\end{align} 
The constant $C$ ($C'$) is determined by the number of massless (massive) 
vector and hypermultiplets $n_V, n_H$ ($n_V', n_H'$), respectively,
\begin{align}
C=\frac{\zeta(3)}{32 \pi^2} \left( n_H-n_V-2 \right)\ , \qquad \label{C}
C'=\frac{1}{32 \pi^2} \left( n_H'-n_V'\right)\  .
\end{align} 
In the minimal case $n_H=n_V=0$, only the supergravity multiplet
contributes to the massless 
sector. Note that only hypermultiplets give rise to positive contributions, 
leading to repulsive behaviour at small distances. A local
minimum can be obtained for $C < 0$, $C' > 0$, and therefore
\begin{equation}
n_H < n_V + 2\ , \qquad n_H' > n_V'\  .
\end{equation}
With $n_H'-n_V' = 1 \ldots \mathcal{O}(100)$, as in heterotic orbifolds \cite{nrx08}, this 
yields the parameter range $10^{-2} \lesssim C' \lesssim 1$.

\begin{figure}
\begin{center}
\includegraphics{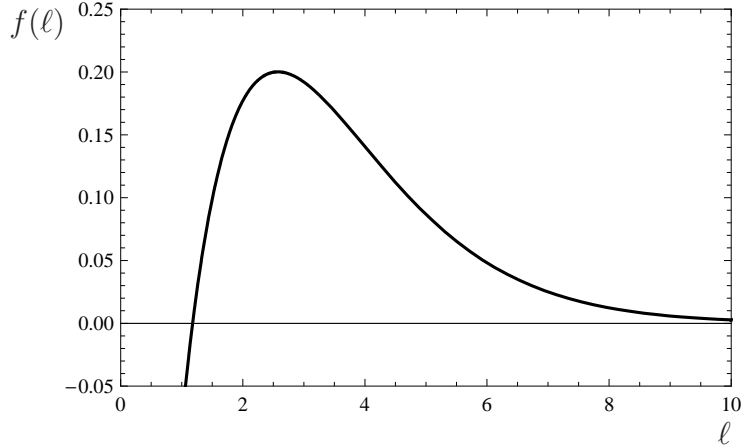} 
\caption{\label{fig:rmass} The function $f(\ell)$, $\ell=LM_\textrm{bulk}$, 
which determines the radion mass $m_{\rho}$ (cf.~(\ref{mrho})).}
\end{center}
\end{figure}

We shall now show how to obtain a ground state with vanishing vacuum energy 
and determine the corresponding compactification scale. For this, we have 
to solve the equations
\begin{equation}\label{conds}
\partial_r V^{(1)}\Big|_{r_0,\sigma_0}=0 \  ,\qquad  
V^{(1)}\Big|_{r_0,\sigma_0}=0\ .
\end{equation}
Imposing $r_0=1,$ we obtain two conditions on the quantities $\ell$ and 
$\sigma_0$, 
\begin{align}
\label{stabscale}\frac{C}{C'}&=\frac{\ell^2}{3}\left[
\frac{\ell}{1-e^\ell}-2\,\textrm{Li}_1(e^{-\ell})\right] - 
\ell\,\textrm{Li}_2(e^{-\ell})-\textrm{Li}_3(e^{-\ell})\  ,\\
\label{uplift}
\frac{\lambda\sigma_0}{C'}& = 
\frac{M_\textrm{bulk}^2}{2 \xi}\left[\frac{\ell}{1-e^\ell} - 
\textrm{Li}_1(e^{-\ell})\right]\ .
\end{align}
The RHS of (\ref{stabscale}) is negative and bounded from below, which 
translates into a condition on the field content (cf.~Fig.~\ref{fig:ratiom}), 
\begin{align}
0<\frac{2-n_H+n_V}{n_H'-n_V'}\lesssim 1.1\ . 
\end{align}
If this bound is satisfied, Eq.~(\ref{stabscale}) can be solved for $\ell$. 
For local minima of the radion potential, this gives the size $L$ in units of $1/M_\textrm{bulk}$ (cf.~Eq.~(\ref{size})). 

Expanding the potential (\ref{pot}) around the local Minkowski vacuum and
using (\ref{uplift}), one obtains for the radion mass  
\begin{equation}\label{mrho}
\frac{m_{\rho}^2}{m_{3/2}^2}=C'\left(\frac{M_\textrm{bulk}}{M}\right)^2 f(\ell)\  ,
\end{equation}
where
\begin{equation}
f(\ell) = 
\frac{2}{3}\left[\frac{\ell\left(1+(\ell-1)e^\ell\right)}{\left(e^\ell-1\right)^2}
- \textrm{Li}_1(e^{-\ell})\right]\  .
\end{equation}
The radion mass vanishes for $\ell=\hat{\ell}\simeq 1.2$, where the ratio 
$C/C'$ is maximal (cf.~Figs.~2,3).
For $\ell>\hat{\ell}$, $m_{\rho}^2$ is positive and we have a stable 
Minkowski vacuum with $L \gtrsim 1/M_{\rm GUT}$ (cf.~(\ref{susymass})).
Fig.~\ref{fig:rmass} also demonstrates that $f(\ell)$ has a local maximum, which
yields an upper bound on the radion mass. For $C'\lesssim 1$, one
obtains 
\begin{equation}
\frac{m_{\rho}^2}{m_{3/2}^2} \lesssim 
0.2\left(\frac{M_\textrm{bulk}}{M}\right)^2\ .
\end{equation}
Larger radion masses require a huge number of massive species.

Having determined the size $L$ of the compact dimension by solving 
Eq.~(\ref{stabscale}), we still have to satisfy Eq.~(\ref{uplift}). This 
is a condition on $\lambda\sigma_0$. Since the RHS of (\ref{uplift}) is 
negative, the coupling $\lambda$ also has to be negative. Given $\lambda$,
this yields a condition on the expectation value $\sigma_0 = X_0\bar{X}_0/M^2$,
and therefore on the parameters of the brane superpotential, which determine
this VEV. This condition represents the fine tuning which is needed to obtain
a Minkowski vacuum. With $M^2_\textrm{bulk}/\xi\simeq 1$, one obtains the upper
bound 
\begin{equation}
|\lambda|\sigma_0 \lesssim 0.4\, C'\  .
\end{equation}
Hence, for $C' < 1$ and $|\lambda|=\mathcal{O} (1)$, the expectation value of $X$ is smaller than the Planck mass. 

\begin{figure}
\begin{center}
\includegraphics{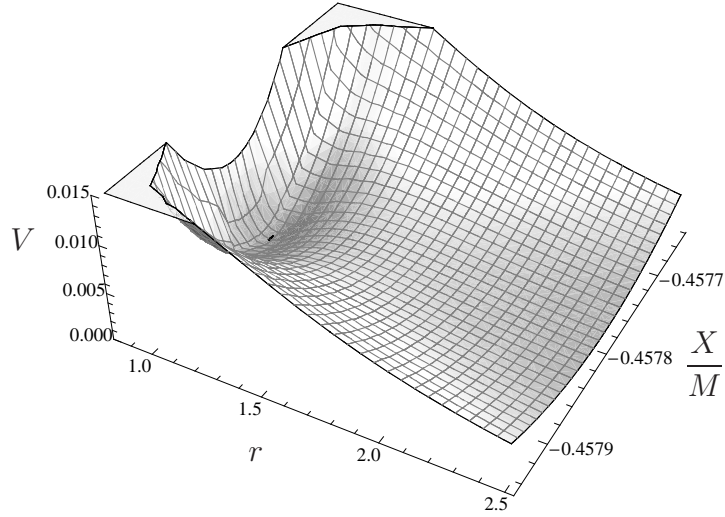} 
\caption{\label{fig:pot3d} The two-field potential $V_F(X,r)$ given by
(\ref{eq:VF2}) in units of $m^2_{3/2}M^2_\textrm{bulk}$, for the choice 
$\xi/M^2_\textrm{bulk} = 1$, $\lambda = -1$, $C'\simeq 1.0$ and 
$\ell\simeq 2.1$.} 
\end{center}
\end{figure}

As an example, consider the superpotential 
\begin{equation}
\label{eq:W}
W(X)=m_{3/2}M^2 \left[2 \frac{X}{\sqrt{\sigma_0}M} + 
\left(\frac{X}{\sqrt{\sigma_0}M}\right)^2\right]\  ,
\end{equation} 
which gives $X=\sqrt{\sigma_0}M$ up to terms 
$\mathcal O(\Delta X/\sqrt{\sigma_0})$ (cf.~\ref{eq:deltaXres}).
Note that Eq.~(\ref{eq:W}) may represent the expansion of a nonperturbative 
brane superpotential up to second order in the field $X$. The corresponding 
two-field potential (\ref{potential}) is given by
\begin{align}
\label{eq:VF2}
V_F(X,r) = & \frac{1}{r^2}\,W_X\bar{W}_{\bar{X}}
- \lambda \xi \frac{\left(X W_X \bar{W} + 
W \bar{X} \bar{W}_{\bar{X}}\right)}{M^4 r^4}
+ V^{(1)}(r,\sigma) + \mathcal O(|W_X|^2 \kappa)\  .
\end{align}
The potential is plotted in Fig.~\ref{fig:pot3d} in the vicinity of $X_0$,
which clearly illustrates the almost no-scale structure of our model 
compared to the no-scale case shown in Fig.~1.

\begin{figure}
\begin{center}
\includegraphics{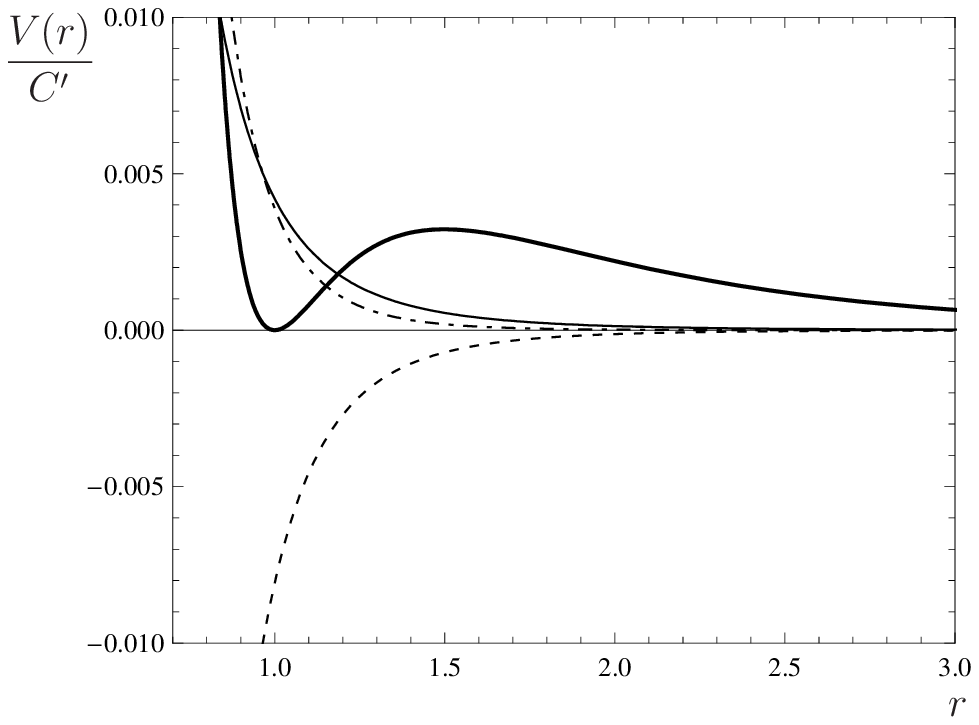} 
\caption{\label{fig:rpot}The radion effective potential $V(r)/C'$ (bold curve),
in units of $m_{3/2}^2 M_\textrm{bulk}^2$ for $\ell\simeq2.1$. The dashed 
(dot-dashed) and the thin curve are the contributions to the Casimir energy 
of massless (massive) bulk degrees of freedom and the FI-term induced  
contribution, respectively, scaled by a factor $1/(100 C')$. } 
\end{center}
\end{figure}

Figure~\ref{fig:rpot} shows the resulting radion potential $V_F(X_0,r)$ for 
$\ell\simeq2.1$, which corresponds to $n_H'-n_V'=2-n_H+n_V$.  
The stable Minkowski vacuum is separated from the runaway solution by a 
barrier of height
\begin{equation}
V_{\rm barrier} \ll m_{3/2}^2 M^2_\textrm{GUT}\  .
\end{equation}
Vanishing of the vacuum energy in the local minimum requires a precise 
cancellation between three different contributions to the potential, all 
$\mathcal{O} (m_{3/2}^2M_{\rm GUT}^2)$. One may also introduce a 
small positive vacuum energy,  $\Lambda \sim (10^{-3}\,  {\rm eV})^4$, which
would correspond to the fine tuning
\begin{equation}
\frac{\Lambda}{m_{3/2}^2M_\textrm{GUT}^2}\sim 10^{-90}\  ,
\end{equation}
for a gravitino mass $m_{3/2}=\mathcal{O} (\rm TeV)$. The rather low
potential barrier implies strong constraints on the maximal temperature
in the early universe \cite{bhx04} as well as the Hubble parameter
during inflation \cite{kl04}.

Stabilization of the compact dimension leaves the pseudoscalar partner of
the radion, an axion, massless. Its coupling to non-Abelian gauge fields
can then lead to a small mass. The contribution from QCD corrections can be 
expressed in terms of the pion mass and decay constant \cite{wei96},
\begin{equation}
m_a \sim \frac{m_{\pi} f_{\pi}}{M} \sim 
10^{-10}\,  \mathrm{eV} \left(\frac{10^{18}\,  \mathrm{GeV}}{M}\right)\ .
\end{equation}
Depending on the cosmological initial conditions and evolution, coherent
oscillations of such ultralight axions (and also the light radion) may yield an unacceptably large
contribution to dark matter.  

It is instructive to compare the described mechanism of radion stabilization with the
approach of \cite{lo03}. In both cases, supersymmetry is
broken by a superpotential localized on a brane, and stabilization is 
achieved by the Casimir energy of massive and massless bulk fields. Here, a
supersymmetric bulk mass 
$M_{\mathrm{bulk}}\simeq M_{\mathrm{GUT}} \gg m_{3/2}$ is 
induced by localized Fayet-Iliopoulos terms \cite{bls07} via the Higgs
mechanism. Finally, the brane field, which provides the non-zero 
superpotential, couples to massive bulk fields. This yields an additional 
contribution to the potential, which has the same order of magnitude as the 
Casimir energy \cite{bcs08}.
In this way, a locally stable Minkowski or de Sitter vacuum can 
be obtained without the need of an additional uplifting mechanism.

\section{Conclusions}

We have considered five-dimensional supergravity theories compactified on the
orbifold $S_1/\mathbb{Z}_2$. The expectation value of a chiral superfield
localized at one of the fixed points, with non-vanishing
superpotential, induces supersymmetry breaking by the radion field. The
result is a no-scale model where the gravitino mass slides with the
undetermined expectation value of the radion field. Perturbative corrections
to the K\"ahler potential, Casimir energy and background values of 
bulk fields induced by localized Fayet-Iliopoulos terms, deform the no-scale
model into an almost no-scale model. The size of the compact dimension is
fixed at $L \sim 1/M_{\mathrm{bulk}} \sim 1/M_{\mathrm{GUT}}$.

Our study of five-dimensional orbifold supergravity models has been motivated
by recent orbifold compactifications of the heterotic string which yield
the supersymmetric standard model in four dimensions \cite{nrx08}, with
orbifold GUTs in five or six dimensions as intermediate step. It will be
interesting to explicitly check whether the one-loop field theory corrections
to the K\"ahler potential considered in this paper are indeed the leading
part of the one-loop string corrections. This may be the case for anisotropic 
orbifold compactifications of the heterotic string leading to orbifold GUTs, 
since the Kaluza-Klein masses which contribute to the Casimir 
energy are smaller than the masses of string excitations. String theory 
also predicts a superpotential for localized chiral superfields and 
couplings of brane to bulk fields. Hence, also radion mediated supersymmetry
breaking may be realized.  

For a general perturbative correction $\kappa$ to the K\"ahler potential,
we have calculated the correction to the effective radion potential to
leading order in $\kappa$. The corresponding radion mass is volume suppressed
compared to the gravitino mass. Moreover, since the stabilization is achieved by quantum corrections, the radion mass is also loop-suppressed.
In addition, a tiny mass for the pseudoscalar partner of the radion, an
axion, is generated by nonperturbative effects of non-Abelian gauge theories.
Hence, the presence of light moduli fields is an unavoidable consequence of
the proposed stabilization mechanism. This is in contrast to models where the
nonperturbative dependence of the superpotential on moduli fields plays
a crucial role. In such models the moduli fields can be heavy 
(cf.~\cite{kkx03,cfx05}).

It remains to be seen whether the light moduli predicted by our stabilization
mechanism are consistent with the various potential `cosmological moduli 
problems'. On the other hand, a radion with a mass two to four orders   
of magnitude smaller than the gravitino mass, could produce a distinctive
signature in the cosmic gamma-ray spectrum and in this way become a `smoking
gun' for the existence of extra dimensions related to the scale of grand
unification.\\

\noindent
{\bf \large Acknowledgement}\\

We would like to thank Laura Covi, Christian Gross, Jan Louis and Kai
Schmidt-Hoberg for helpful discussions.

\end{document}